\documentclass[aps,superscriptaddress,preprintnumbers,amsmath,amssymb,a4paper,preprint]{revtex4}
\usepackage{graphicx}
\usepackage{dcolumn}
\usepackage{bm}
\usepackage{color}

\begin{document}

\title{Emitter--site selective photoelectron circular dichroism of trifluoromethyloxirane}

\author{\firstname{M.} \surname{Ilchen}}\email{ilchen@slac.stanford.edu}
\affiliation{SLAC National Accelerator Laboratory, 2575 Sand Hill Road, Menlo Park, California 94025, USA}
\affiliation{Stanford PULSE Institute, 2575 Sand Hill Road, Menlo Park, California 94025, USA}
\affiliation{European XFEL GmbH, Holzkoppel 4, 22869 Schenefeld, Germany}

\author{\firstname{G.} \surname{Hartmann}}
\affiliation{Institut f\"{u}r Physik und CINSaT, Universit\"{a}t Kassel, Heinrich-Plett-Str. 40, 34132 Kassel, Germany}
\affiliation{Deutsches Elektronen-Synchrotron DESY, Notkestra{\ss}e 85, 22607 Hamburg, Germany}

\author{\firstname{P.} \surname{Rupprecht}}
\altaffiliation{Present address: Max-Planck-Institut f\"{u}r Kernphysik, Saupfercheckweg 1, 69117 Heidelberg, Germany}
\affiliation{SLAC National Accelerator Laboratory, 2575 Sand Hill Road, Menlo Park, California 94025, USA}
\affiliation{Stanford PULSE Institute, 2575 Sand Hill Road, Menlo Park, California 94025, USA}
\affiliation{Physik-Department, Technische Universit\"{a}t M\"{u}nchen, James-Franck-Strasse 1, 85748 Garching, Germany}

\author{\firstname{A.\,N.} \surname{Artemyev}}
\affiliation{Institut f\"{u}r Physik und CINSaT, Universit\"{a}t Kassel, Heinrich-Plett-Str. 40, 34132 Kassel, Germany}

\author{\firstname{R.\,N.} \surname{Coffee}}
\affiliation{SLAC National Accelerator Laboratory, 2575 Sand Hill Road, Menlo Park, California 94025, USA}

\author{\firstname{\\Z.} \surname{Li}}
\affiliation{SLAC National Accelerator Laboratory, 2575 Sand Hill Road, Menlo Park, California 94025, USA}
\affiliation{Center for Free-Electron Laser Science CFEL, DESY, Notkestra{\ss}e 85, 22607 Hamburg, Germany}

\author{\firstname{H.} \surname{Ohldag}}
\affiliation{SLAC National Accelerator Laboratory, 2575 Sand Hill Road, Menlo Park, California 94025, USA}

\author{\firstname{H.} \surname{Ogasawara}}
\affiliation{SLAC National Accelerator Laboratory, 2575 Sand Hill Road, Menlo Park, California 94025, USA}

\author{\firstname{T.} \surname{Osipov}}
\affiliation{SLAC National Accelerator Laboratory, 2575 Sand Hill Road, Menlo Park, California 94025, USA}

\author{\firstname{D.} \surname{Ray}}
\affiliation{SLAC National Accelerator Laboratory, 2575 Sand Hill Road, Menlo Park, California 94025, USA}

\author{\firstname{Ph.} \surname{Schmidt}}
\affiliation{Institut f\"{u}r Physik und CINSaT, Universit\"{a}t Kassel, Heinrich-Plett-Str. 40, 34132 Kassel, Germany}

\author{\firstname{\\T.\,J.\,A.} \surname{Wolf}}
\affiliation{Stanford PULSE Institute, 2575 Sand Hill Road, Menlo Park, California 94025, USA}

\author{\firstname{A.} \surname{Ehresmann}}
\affiliation{Institut f\"{u}r Physik und CINSaT, Universit\"{a}t Kassel, Heinrich-Plett-Str. 40, 34132 Kassel, Germany}

\author{\firstname{S.} \surname{Moeller}}
\affiliation{SLAC National Accelerator Laboratory, 2575 Sand Hill Road, Menlo Park, California 94025, USA}

\author{\firstname{A.} \surname{Knie}}
\affiliation{Institut f\"{u}r Physik und CINSaT, Universit\"{a}t Kassel, Heinrich-Plett-Str. 40, 34132 Kassel, Germany}

\author{\firstname{Ph.\,V.} \surname{Demekhin}}\email{demekhin@physik.uni-kassel.de}
\affiliation{Institut f\"{u}r Physik und CINSaT, Universit\"{a}t Kassel, Heinrich-Plett-Str. 40, 34132 Kassel, Germany}
\affiliation{Research Institute of Physics, Southern Federal University, Stachki\,av.\,194, 344090 Rostov-on-Don, Russia}

\date{\today}

\begin{abstract}
The angle-resolved inner-shell photoionization of R-trifluoromethyloxirane, C$_3$H$_3$F$_3$O,  is studied experimentally and theoretically. Thereby, we investigate the photoelectron circular dichroism (PECD) for nearly-symmetric O 1s and F 1s  electronic orbitals, which are localized on different molecular sites. The respective dichroic $\beta_1$ and angular distribution $\beta_2$  parameters are measured at the  photoelectron kinetic energies from 1 to 16 eV by using variably polarized synchrotron radiation and velocity map imaging spectroscopy. The present experimental results are in good agreement with the outcome of \emph{ab initio} electronic structure calculations. We report a sizable chiral asymmetry $\beta_1$ of up to about 9\% for the K-shell photoionization of oxygen atom. For the individual fluorine atoms, the present calculations predict asymmetries of similar size. However, being averaged over all fluorine atoms, it drops down to about 2\%, as also observed in the present experiment. Our study demonstrates a strong emitter- and site-sensitivity of PECD in the one-photon inner-shell ionization of this chiral molecule.
\end{abstract}


\maketitle

\section{Introduction}
\label{sec:intro}

In 1976, Ritchie has predicted theoretically that angle-resolved photoelectron spectra of chiral molecules exhibit a sizable circular dichroism (CD) effect, which in contrast to the normal CD in total absorption spectra is governed by the electric dipole interaction \cite{Ritchie76}. It took about 25 years to verify these predictions experimentally \cite{Boewerling01}. Since then, the photoelectron circular dichroism (PECD) in the one-photon ionization of chiral molecules in the gas phase has been extensively studied experimentally and theoretically. At present, numerous of independent works have confirmed that for randomly-oriented molecules, PECD can be seen as a forward/backward asymmetry in the emission of photoelectrons which is typically on the order of a few per cents. Most of those studies are reviewed in Refs.~\cite{Powis08b,Nahon10b,Nahon15}.

The first experiments performed with circularly polarized synchrotron radiation on randomly oriented bromocamphor \cite{Boewerling01} and camphor \cite{Garcia03} illustrated a sizable PECD of about 3--4\%. It was also noticed that PECD changes with the binding energy of a system, i.e., it is different for the ionization of different molecular orbitals \cite{Boewerling01}. Furthermore, a strong dependence of PECD on the photoelectron kinetic energy (exciting-photon energy) was found experimentally and theoretically for outer-shell photoionization of methyloxirane \cite{Turchini04,Stranges05,Garcia13}, chiral derivates of oxirane \cite{Stener04}, as well as  camphor and fenchone \cite{Nahon06,Powis08a,Nahon16}.

Nowadays, PECD in the one-photon ionization of outer electrons is a well-established research area, which includes studies of  molecular dimers \cite{Nahon10a}, clusters \cite{Daly11}, metal-organic complexes \cite{Catone12}, and even small biological molecules \cite{Giardini05,Turchini09,Tia13,Tia14}. Experimentally, this effect has been studied by tunable circularly polarized synchrotron radiation utilizing different methods of angle-resolved photoelectron spectroscopy. Theoretically, the Continuum Multiple Scattering method with the local $X\alpha$ exchange correlation (CMS-$X\alpha$) \cite{MSXalp} and the time-dependent density functional theory (TDDFT) B-spline LCAO formalism \cite{TDDFT} were used in those studies.

Molecular orbitals of the outer valence electrons (HOMO$-n$) are typically delocalized over a large part of a molecule and are strongly asymmetric. This asymmetry of the initial electronic state is naturally imprinted in the observed PECD through the photoionization  amplitudes. Notwithstanding, there is another contribution to PECD, which is related to the final electronic state through the same amplitudes. Indeed, a contrastive theoretical study of chiral derivates of oxirane \cite{Stener04} has suggested that the magnitude of PECD is also governed by the ability of the outgoing photoelectron continuum wave to probe the asymmetry of the molecular ion potential.

The latter effect of the final electron continuum state plays a decisive role in the photoionization of almost-symmetric inner-shell electrons. The very first study of inner-shell PECD performed for the O=C(1s) ionization of camphor \cite{Hergenhahn04} reported a large chiral asymmetry, which was shown to be dependent on the photoelectron kinetic energy scanned up to 65 eV above the ionization threshold. These experimental results were supported by numerical simulations, which suggested that PECD is caused here by the final-state scattering effects on the chiral potential of a molecule \cite{Hergenhahn04}.

At present, there are several works reporting  PECD after inner-shell photoionization of chiral molecules \cite{Stener04,Nahon06,Powis08a,Hergenhahn04,Alberti08,Ulrich08,Tia16}. As an advantage, it allows  to selectively address electronic orbitals localized on a particular chiral center of a molecule. Importantly, inner-shell ionization, followed by an ultrafast Auger decay and subsequent fragmentation of the resulting dication by Coulomb explosion, offers the possibility to access molecular frame photoelectron angular distributions by multicoincident detection techniques \cite{Jahnke02}. This principle has recently been utilized to study PECD in the O 1s photoionization of uniaxially oriented methyloxirane \cite{Tia16}.

Very recently \cite{Nahon15,Garcia17}, PECD in the one-photon ionization of electrons from the HOMO and HOMO-1 of trifluoromethyloxirane  (C$_3$H$_3$F$_3$O) has been studied in the vibriationally-resolved mode. For this chiral molecule, PECD after inner-shell photoionization has not yet been  investigated. In order to study the PECD at different emitter sites of this molecule, we address in this work the O 1s and F 1s photoionization of R-trifluoromethyloxirane. Our experimental and theoretical methods are described in Sec.~\ref{sec:EXPTHE}. The measured and calculated dichroic and angular distribution parameters are compared and discussed in Sec.~\ref{sec:results}. We conclude in Sec.~\ref{sec:concl} with a brief summary and outlook.

\section{Experiment and theory}
\label{sec:EXPTHE}

The present experiments were performed at the Stanford Synchrotron Radiation Lightsource (SSRL) at the BL13-2 beamline \cite{BEAMLINE,Katayama2013}. Equipped with the 26-pole elliptically polarizing undulator \cite{UNDUL}, this synchrotron beamline can produce  $94\pm 6\%$ circularly polarized soft X-rays. The spherical grating monochromator at BL13-2 can achieve a spectral resolving power of $10^4$ over an energy range from 180 to 1100 eV with a flux of $10^{11}$ to $10^{12}$ photons/sec.

The sample, (R)-(+)-3,3,3-Trifluoro-1,2-epoxypropane (97\%, Sigma Aldrich), was introduced into the interaction chamber by an effusive gas jet via a 0.5\,mm skimmer to reach the vacuum conditions necessary for the beamline operation. The interaction region of gas jet and synchrotron radiation coincided with the focus point of the velocity map imaging (VMI) spectrometer of the LAMP chamber of the LCLS AMO beamline \cite{Osipov12}. The electric fields were optimized to collect electrons with kinetic energies of around 10~eV with an energy resolution of about 5\%.  The acquired data sets for the O and F K-edges cover photoelectron kinetic energies up to 16~eV while using circularly (with both positive `+' and negative `--' helicities) and  linearly polarized light. In order to allow for detector calibration as well as for an estimate of the degree of linear/circular polarization,  an additional racemic mixture of the target was  introduced into the interaction chamber.

An inverse Abel transformation was applied to reconstruct slices of the 3D photoelectron angular distributions out of the 2D-VMI images. Several transformation methods, namely the BASEX-method \cite{Dribinski2002},  direct integration of the Abel integral  \cite{Dribinski2002}, an iterative approach \cite{Vrakking2001}, the lin-BASEX \cite{Gerber2013}, the Hansen-Law-algorithms \cite{Hansen1985}, and onion-peeling \cite{Dasch1992,Bordas1998,Rallis2014}  were used as an evaluation stability proof and as a means to estimate the uncertainty. The dichroic parameter $\beta_1$ and angular distribution parameter $\beta_2$ were determined independently. The difference between two normalized photoelectron angular distributions measured with the circularly polarized light was fitted to extract  $\beta_1$ using the relation
$I_+(\theta)-I_-(\theta)=2\beta_1P_1(\cos \theta)$. The normalized photoelectron angular distribution measured with the linearly polarized light reveals  $\beta_2$ via the parametrization $I(\theta)=1+\beta_2 P_2(\cos \theta)$. Uncertainties are estimated by statistics error perturbation, fit quality, and 3D-reconstruction stability.

The laboratory-frame angular distribution of the photoelectrons emitted by randomly-oriented molecules excited by circularly polarized light is described by the well-known formula for the differential photoionization cross section \cite{Ritchie76,Cherepkov81molspinpol,Cherepkov82chiral}:
\begin{equation}
\frac{\mathrm{d}\sigma^\pm}{\mathrm{d\Omega}}=\frac{ {\sigma} }{4\pi}\left[ 1 \pm \beta_1 P_1(\cos \theta ) -\frac{1}{2} \beta_2 P_2(\cos \theta )\right]. \label{eq:pi_dcs}
\end{equation}
Here, `$\pm$' stands for the positive and negative helicity of the circularly polarized radiation, $P_L(\cos \theta )$  are  the Legendre polynomials, and $\theta$ is the angle between the direction of the propagation of the exciting-radiation and the direction of the emission of photoelectrons encompassed in the solid angle $d\Omega$. The total photoionization cross section $\sigma$ and parameters $\beta_{L}~{(L=1,2)}$ can be computed via the following equations \cite{Demekhin0910,Knie14}
\begin{subequations}
\begin{equation}
{\sigma} =  \sum_{\Lambda_0\Lambda_1}  \sum_{\ell m  q  }  \vert \langle\Lambda_1  \varepsilon\ell m    \vert \textbf{d}_q\vert \Lambda_0\rangle \vert^2,\label{eq:pi_tcs}
\end{equation}
\vspace{-0.5cm}
\begin{multline}
\beta_L = \frac{1}{{\sigma}} \sum_{\Lambda_0\Lambda_1} \sum_{\ell m q  } \sum_{\ell^\prime m^\prime q^\prime} (i)^{\ell+\ell^\prime} (-1)^{\ell^\prime+m+q}  \\ \times  e^{i(\delta_{\ell m }-\delta_{\ell^\prime m^\prime })} \sqrt{\frac{3L(8+L)}{2}(2\ell+1)(2\ell^\prime+1)}\\ \times  \left(\begin{array}{ccc}\ell &~~\ell^\prime&L\\ -m& m^\prime & q- q^\prime \end{array}\right) \left(\begin{array}{ccc}~~1 &~1&L\\ ~q  & -q^\prime& ~m^\prime-m \end{array}\right)   \\ \times\left(\begin{array}{ccc}\ell &\ell^\prime&L\\ 0&0&0 \end{array}\right)
\langle\Lambda_1  \varepsilon\ell m    \vert \textbf{d}_q\vert \Lambda_0\rangle \langle\Lambda_1  \varepsilon\ell^\prime m^\prime   \vert \textbf{d}_{q^\prime}\vert \Lambda_0\rangle^\ast. \label{eq:bel}
\end{multline}
\end{subequations}
In these equations, $\langle\Lambda_1  \varepsilon\ell m    \vert \textbf{d}_q\vert \Lambda_0\rangle$ is the dipole transition matrix element for the ionization of the electronic state $\Lambda_0$ by the linearly ($q=0$) or circularly ($q=\pm1$) polarized light, which results in the population of the final ionic state $\Lambda_1$ and emission of the photoelectron partial wave $  \varepsilon\ell m  $ with the kinetic energy $\varepsilon$, angular momentum  $\ell$  and its projection $m$ (quantum numbers $m$ and $q$ are defined in the molecular quantization frame). Further on, $\delta_{\ell m }$ is the  phase shift of the partial electron wave, and summations over indices $\Lambda_{0,1}$  must be performed over all degenerate electronic states.

The photoionization transition amplitudes were calculated in the present work by the Single Center (SC) method and code \cite{Demekhin11,Galitskiy15}, which provides an accurate description of the partial photoelectron continuum waves in molecules. It was already successfully applied to study angle-resolved spectra of diatomic molecules \cite{Demekhin0910,Demekin10a,Demekin10b,Demekin10c}, weakly-bound dimers \cite{Sann16}, polyatomic molecules \cite{Knie14,Antonsson15,Knie16}, small metallic clusters \cite{Galitskiy15}, and even chiral molecules \cite{Tia16,Artemyev15}. Briefly,  a  molecular orbital is represented in the SC method with respect to a single center of the molecule via an expansion in terms of spherical harmonics. The wave functions of a photoelectron in the continuum are sought  as  numerical solutions of the system of coupled one-particle equations for the radial partial electron waves with accurate molecular field potentials, provided by all nuclei and  occupied electronic shells of a molecule. Thereby, the main physical mechanism for handing a chiral asymmetry of the molecular ion potential over to the outgoing photoelectron wave by multiple scattering effects is accurately described by the method. The numerical scheme implies non-iterative accounting for the non-local exchange Coulomb interaction of a photoelectron with a molecular core, which makes the computational procedure robust \cite{Demekhin11,Galitskiy15}.

In the present work, photoionization transition amplitudes were computed in the frozen-core Hartree-Fock approximation at the equilibrium internuclear geometry of the ground electronic state of R-trifluoromethyloxirane, optimized at the (2,2)-CASSCF/6-31G(d,p) level of theory. Molecular orbitals of the occupied electronic shells, computed by the PC GAMESS (General Atomic and Molecular Electronic Structure System,  \cite{Schmidt93}), version Alex A. Granovsky \cite{GAMESS}, were represented relatively to the nuclear charge center of gravity (chosen as the molecular center) by expansions over spherical harmonics with $\ell,\vert m \vert  <60$. The SC expansions of the continuum partial waves were restricted by  partial harmonics with $\ell_\varepsilon, \vert m_\varepsilon \vert < 35$. Finally, multipole expansions of the exchange Coulomb interaction of the photoelectron with the core electrons \cite{Galitskiy15} were restricted by harmonics with $k, \vert q \vert < 15$.

\section{Results and discussion}
\label{sec:results}

\begin{figure}
\includegraphics[scale=0.425]{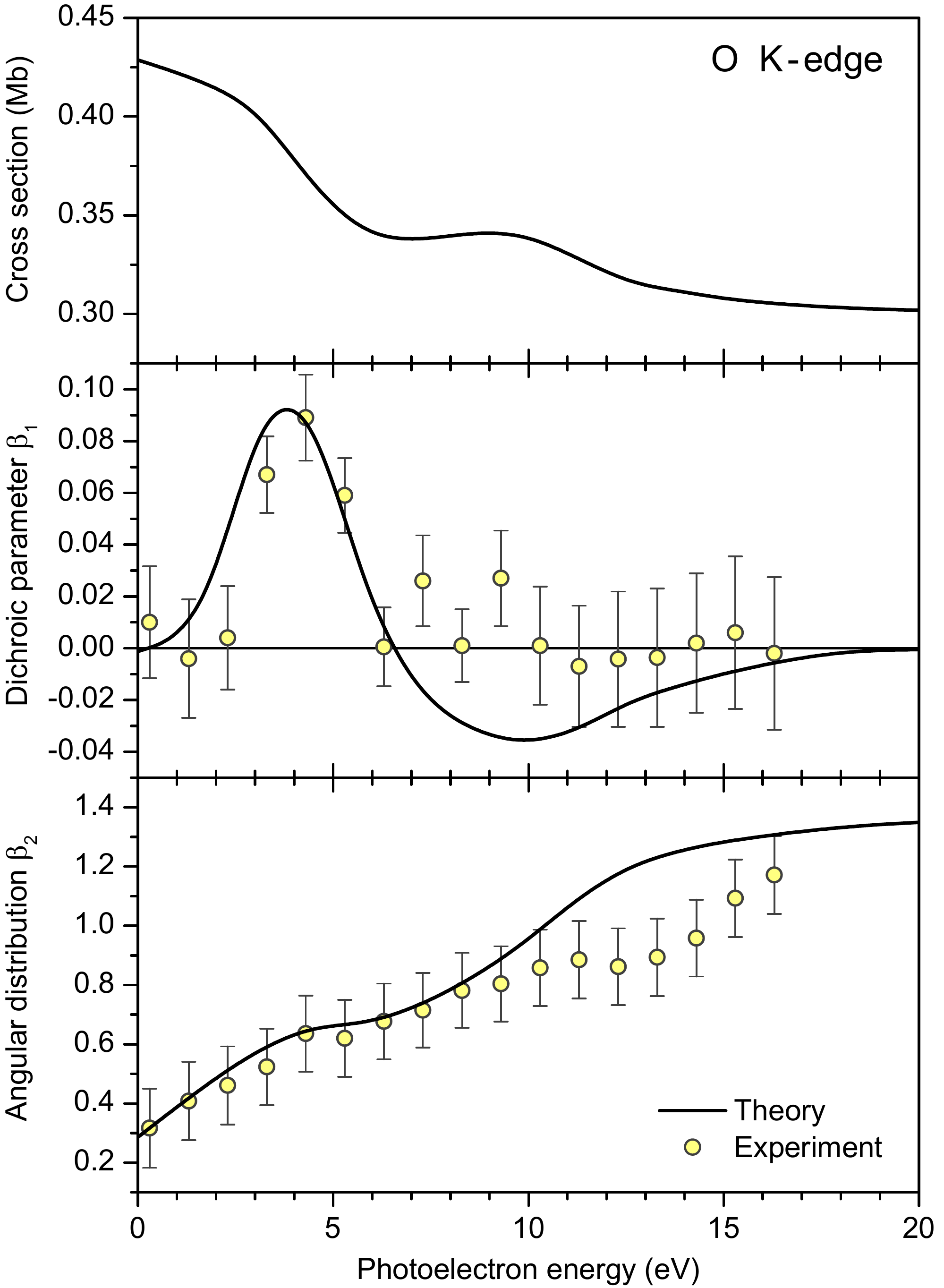}
\caption{(Color online) The presently measured (circles with error bars) and computed (solid curves) cross section (uppermost panel), dichroic parameter $\beta_1$ (middle panel), and angular distribution parameter $\beta_2$ (lowermost panel) for the O 1s photoionization of R-trifluoromethyloxirane as functions of the photoelectron energy.}\label{fig_O1s}
\end{figure}

\begin{figure}
\includegraphics[scale=0.425]{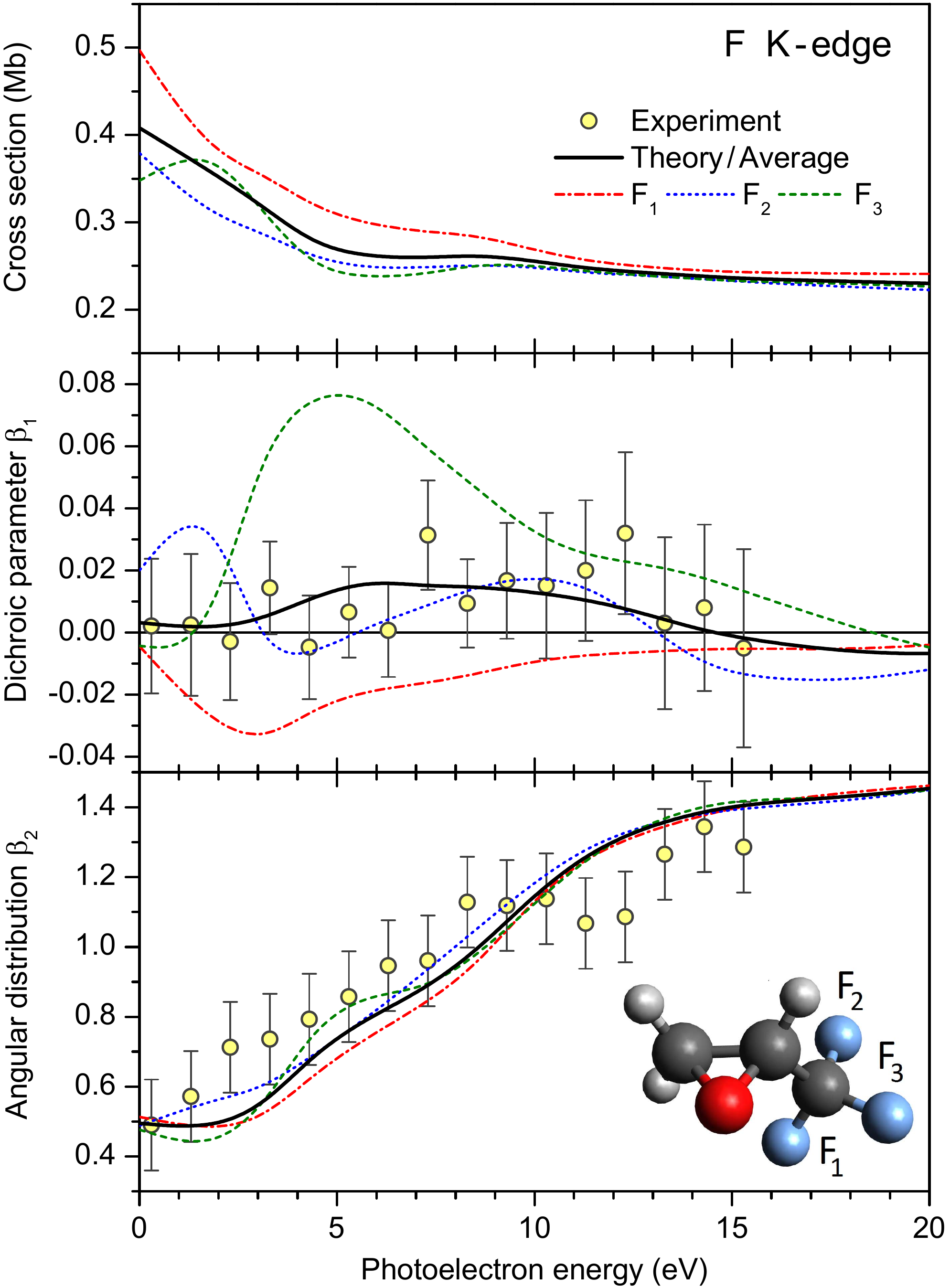}
\caption{(Color online) The presently measured (circles with error bars) and computed (curves) cross sections (uppermost panel), dichroic parameters $\beta_1$ (middle panel), and angular distribution parameters $\beta_2$ (lowermost panel) for the F 1s photoionization of R-trifluoromethyloxirane as functions of the photoelectron energy. Results computed for the individual fluorine atoms $F_i$, as enumerated in the inset in the lowermost panel, are shown by broken curves, whereas the final data, averaged over three atoms, are drawn as solid curves.}\label{fig_F1s}
\end{figure}

The present angle-resolved  spectra for the K-shell photoionization of oxygen  and fluorine atoms are collected  in Figs.~\ref{fig_O1s} and \ref{fig_F1s}, respectively. These figures depict  (from the top to bottom) the total cross section $\sigma$, dichroic parameter $\beta_1$, and angular distribution parameter $\beta_2$ computed as functions of the electron kinetic energy up to 20 eV. For the photoelectron energies up to 16~eV, the computed  $\beta_1$ and $\beta_2$ parameters are compared with the presently measured respective values. Determining total cross sections experimentally (even on the arbitrary scale) requires full characterization of the light source and detection apparatus in order to enable interrelation of the total electron signals recorded at different exciting-photon energies, which is beyond the scope of the present work. We thus do not report experimental $\sigma$ values here.

For higher photoelectron energies from 10 to 16 eV, the recorded velocity map images were partially compromised in the forward direction. Therefore, only part of the data was taken into account for the $\beta_1$  determination. The procedure of reconstructing data with a reduced image area and, therefore, reduced statistical validity was carefully cross checked with different transformation methods \cite{Dribinski2002,Vrakking2001,Gerber2013,Hansen1985,Dasch1992,Bordas1998,Rallis2014} and is taken into account in the determination of uncertainties, which can be clearly seen in the middle panels of Figs.~\ref{fig_O1s} and \ref{fig_F1s} in the energy range of 10--16~eV. This problem did not affect the accuracy of the measured $\beta_2$ parameter, since the main dipole contributions of the electron signal, which are important for the determination of positive  $\beta_2$ values, were always in the detection area.

As discussed above, dichroic parameters were determined from the difference between the two normalized photoelectron signals corresponding to the use of two opposite circular polarizations of exciting synchrotron radiation. This procedure automatically implies a partial compensation of the extrinsic signals deposited in each spectrum as a symmetric background (noise). On the contrary, this noise cannot be excluded from the data analysis when determining angular distribution parameter out of the single spectrum acquired with the linearly polarized radiation. This led to larger error bars for the measured $\beta_2$ values compared to those for $\beta_1$ (see lowermost panels of Figs.~\ref{fig_O1s} and \ref{fig_F1s}). Finally, the one-particle frozen-core Hartree-Fock approximation, used in the present calculations, does not include electron correlations and core relaxation effects. Neglecting these effects may result in an underestimation of the  absolute values of the total photoionization cross section very close to the ionization threshold by up to a factor of 2 \cite{Sukhorukov87,Sukhorukov91,Arp93}.

The middle panel of Fig.~\ref{fig_O1s} demonstrates an overall good agreement between the $\beta_1$  values computed and measured  for the O K-edge. In most of the cases the  deviation between  theory and experiment is almost within the experimental error bars. Both, theory and experiment indicate a sizable chiral asymmetry of $\beta_1 \approx  9\%$ around $\varepsilon=4$~eV. From the lowermost panel of Fig.~\ref{fig_O1s} one can see, that the presently computed and measured angular distribution parameters $\beta_2$ agree  well, and they start to deviate from each other only for the electron energies above about 10~eV. This might be related to the fact that the presently measured $\beta_2$ values are slightly underestimated due to the presence of a symmetric background in the collected electron signal as discussed above.

The $\sigma$,  $\beta_1$, and  $\beta_2$ parameters computed for the individual fluorine atoms are depicted in Fig.~\ref{fig_F1s} by broken curves. The numbering of $F_i$ atoms is indicated in the inset at the  bottom of this figure. As one can see, the cross sections  $\sigma$ and angular distribution parameters $\beta_2$, computed for different atoms, are very similar for higher photoelectron energies, and they differ from each other only for electron energies below about 10~eV. On the contrary, dichroic parameters $\beta_1$ of the individual fluorines (middle panel) are very different from each other: $\beta_1(F_1)$ is mainly negative; $\beta_1(F_3)$ is mainly positive; whereas $\beta_1(F_2)$ changes its sign as a function of electron kinetic energy. The present theory predicts individual asymmetry to be on the order of about  $\beta_1 \approx 8\%$ for the $F_3$ atom.

In the present experiment, the 1s photoelectrons emitted from different fluorine atoms were not resolved. In order to compare the present theoretical and experimental data, the computed $\beta_L$-parameters need to be averaged over three fluorine atoms as follows:
\begin{subequations}
\begin{equation}
{\sigma} =\frac{1}{3} \sigma_{tot} = \frac{1}{3}\sum_{i=1}^3\sigma(F_i),
\label{eq:CS_aw}
\end{equation}
\vspace{-0.5cm}
\begin{equation}
\beta_L=\frac{1}{\sigma_{tot}}\sum_{i=1}^3\sigma(F_i)\beta_L(F_i). \label{eq:bel}
\end{equation}
\end{subequations}
The average theoretical parameters  $\sigma$,  $\beta_1$, and  $\beta_2$  are depicted in Fig.~\ref{fig_F1s} by solid curves. Note that the average cross section $\sigma$ in Eq.~(\ref{eq:CS_aw}) and in the uppermost panel of Fig.~\ref{fig_F1s} differs from the total cross section $\sigma_{tot}$  by a factor of 3. As a result of this averaging, the computed chiral asymmetry drops significantly down, and it does not exceed now 2\%. This is consistent with the present observation, as can be seen from the middle panel of Fig.~\ref{fig_F1s} which demonstrate a very good agreement between the average theoretical and experimental $\beta_1$ parameters. Similarly to the case of the O K-edge, the averaged computed and the measured angular distribution parameters $\beta_2$ for the F K-edge agree very well for the lower electron energies, and start to deviate for the energies above  10~eV.

\section{Conclusion}
\label{sec:concl}

Dichroic parameters $\beta_1$ and angular distribution parameters $\beta_2$ for the 1s photoionization of  O and F atoms in R-trifluoromethyloxirane were measured and computed for different electron kinetic energies above the respective ionization thresholds. The experiment was performed at the BL13-2 beamline of SSRL (SLAC) utilizing variably polarized soft X-rays and velocity map imaging spectroscopy. Electronic structure calculations were carried out by the Single Center method in the frozen-core Hartree-Fock approximation.

The present calculations demonstrate strong dispersions of the dichroic parameters $\beta_1$  for O and individual F atoms in trifluoromethyloxirane, which for some photoelectron energies reach about 9\%. For the oxygen K-edge, this theoretical result is in full agreement with the experiment. In order to compare theoretical and experimental results for the fluorine K-edge, the computed data were additionally averaged over the three F atoms. This results in  a considerable drop of the maximal value of $\beta_1$ down to about 2\% which also agrees with the present observations.

The present study provides opportunities for future investigations of the molecular frame \cite{Tia16} photoelectron circular dichroism in trifluoromethyloxirane. As the next step, we plan to  extend it towards exploration of transient chirality accompanying fragmentation dynamics by inner-shell pump -- inner-shell probe experiments at X-ray free-electron lasers \cite{Lutman16,Hartmann16}.

\begin{acknowledgements}
The experimental part of this work was carried out in the frame of a collaboration agreement between LCLS and SSRL at SLAC (USA) under contract number DE-AC02-76SF00515 of the U.S. Department of Energy, Office of Science, Office of Basic Energy Sciences. This work was partly supported by the State Hesse excellence initiative LOEWE within the focus-project ELCH, by the Deutsche Forschungsgemeinschaft (DFG project No. DE 2366/1-1), and by the Russian Foundation for Basic Research (grant no. 16-03-00771a). M.I. and Z.L. acknowledge funding from the Volkswagen foundation within the Peter Paul Ewald-Fellowship. P.R. acknowledges funding from the German Academic Exchange Service (DAAD). T.J.A.W. thanks the German National Academy of Sciences Leopoldina for a fellowship (Grant No. LPDS2013-14). Ph.V.D. acknowledges Research Institute of Physics Federal University for the hospitality and support during his research stay there.
\end{acknowledgements}

\end{document}